\documentclass[a4paper,aps,prl,groupedaddress,showkeys,
  superscriptaddress,floats,floatats,floatfix]{revtex4}
\usepackage[latin1]{inputenc}
\usepackage{graphicx}  
\usepackage{dcolumn}   
\usepackage{bm}        
\usepackage{natbib}
\usepackage{latexsym}
\usepackage{mathrsfs}
\usepackage{amssymb}
\usepackage{amsmath}
\usepackage{amscd}
\usepackage{color}
\usepackage{verbatim}
\begin{document}
\title{Ratchetlike pulse controlling the Fermi deceleration and
  hyperacceleration}  
\author{Cesar Manchein}
\email[E-mail address:~]{cmanchein@gmail.com}
\affiliation{Departamento de F\'\i sica, Universidade Federal do Paran\'a,
         81531-990 Curitiba, PR, Brazil}
\author{Marcus W. Beims}
\email[E-mail address:~]{mbeims@fisica.ufpr.br}
\affiliation{Departamento de F\'\i sica, Universidade Federal do Paran\'a,
         81531-990 Curitiba, PR, Brazil}


\begin{abstract}
Using an ac driven asymmetric pulse we show how the Fermi acceleration
(deceleration) can be controlled. A {\it deformed} sawtooth (Ratchetlike)
pulse representing the moving wall in the static Fermi-Ulam model is
considered. The time integral from the pulse over one period of oscillation 
must be negative to obtain deceleration and positive to obtain hyperacceleration.
We show that while the decelerated case is chaotic, for the hyperaccelerated
case the Lyapunov exponents converge to zero. Numerical simulations indicate that
the hyperaccelerated case is ergodic in velocity space. Switching between 
different pulse deformations we are able to control the particle acceleration. 
Results should be valid for any pulse for which the time integral can be 
manipulated between positive and negative values. 
\end{abstract}



\maketitle

Many years ago Enrico Fermi~\cite{fermi49} suggested an acceleration mechanism
of cosmic ray particles interacting with a time dependent magnetic field. Later
on different versions of the original model for the Fermi acceleration were 
proposed. In the first one, the Fermi-Ulam (FU) model, a bouncing particle
moves between a fixed surface and a parallel oscillating surface \cite{ulam61}. 
This model was shown to be chaotic \cite{lichtenberg72,lichtenberg92}. In order
to improve simulations a simplified version of the FU model was proposed
\cite{lichtenberg72}, called the static wall model. It ignores the displacement 
of the moving wall but keeps the essential information for the momentum transfer 
as the wall was oscillating. This static model was discussed in many aspects 
\cite{lichtenberg72,leonel04,leonel04-prl,leonel05}, even for circular billiards 
\cite{egydio06}. Usually invariant curves in the phase space, found for
higher velocities, prevent the particle to increase its kinetic energy without 
bounds. Recently the hopping wall 
approximation was proposed \cite{karlis06,karlis07} which takes into
account the effect of the wall displacement, and allows the analytical 
estimation of the particle mean velocity. Compared to the simplified
static model, the particle acceleration is enhanced. The second kind of Fermi
accelerated model was proposed in 1977 by Pustyl'nikov 
\cite{pustylnikov77}, who considered a particle on a periodically 
oscillating horizontal surface in the presence of a gravitational field. 
The above topics got attention in various areas of physics, ranging 
from nonlinear physics
\cite{lichtenberg72,lichtenberg92,leonel04,leonel04-prl,leonel05,egydio06,
karlis06,karlis07}, atom optics \cite{steane95,schleich98,saif05}, plasma 
physics \cite{milovanov01,michalek99} to astrophysics 
\cite{veltri04,kobayakawa02,malkov98}.

In this paper we use an ac driven asymmetric pulse to control the acceleration 
(deceleration) in the simplified FU model. The pulse is a {\it deformed} 
sawtooth driving law for the moving wall. This Ratchetlike pulse differs from 
the ac driven asymmetric pulses ({\it symmetric} 
sawtooth) used for the Fermi acceleration in the early work of Lichtenberg 
{\it et al.}~\cite{lichtenberg72} and proposed recently to control the motion of 
magnetic flux quanta \cite{cole06} and to analyze the relative  efficiency of 
mechanism leading to increased acceleration in the hopping wall approximation 
\cite{karlis07}. In the simplified Fermi model~\cite{lichtenberg72} the
particle is free to move between the elastic impacts with the walls. Consider 
that the moving wall oscillates between two extrema with amplitude $v_0$.
The gravitational force is considered zero. The system is described by a map 
$M_{1(2)}(V_n,\phi_n)=(V_{n+1},\phi_{n+1})$ which gives, respectively, the 
velocity of the particle, and the phase of the moving wall, immediately after 
the particle suffers a collision with the wall.  Considering dimensionless
variables the FU map with the {\it deformed} sawtooth wall is written as 

\begin{equation}
M_1:\left\{
\begin{array}{ll}
  V_{n+1} = \left\vert V_{n}+\dfrac{v_0}{\eta_1}(\phi_n-\eta_1) \right\vert, \\
  \phi_{n+1} = \phi_{n} + \mu\dfrac{(\eta_1+\eta_2)}{V_{n+1}} 
\hspace{7mm} \mbox{mod}~(\eta_1+\eta_2),  
\end{array}
\right. 
\label{map_1}
\end{equation}
for $\phi_n<\eta_1$, and
\begin{equation}
M_2:\left\{
\begin{array}{ll}
  V_{n+1} = \left\vert V_{n}+\dfrac{v_0}{\eta_2}(\phi_n-\eta_1) \right\vert, \\
  \phi_{n+1} = \phi_{n} + \mu\dfrac{(\eta_1+\eta_2)}{V_{n+1}} \hspace{7mm}
  \mbox{mod}~(\eta_1+\eta_2), 
\end{array}
\right. 
\label{map_2}
\end{equation}
for $\phi_n\ge\eta_1$, 
where $n$ is the iteration number and $\mu$ is the maximum distance between the 
walls. Since in this simplified model the displacement of the moving wall is
ignored, the modulus function is used to avoid errors due to successive
collisions which may occur in the original model. In other words, if after 
a collision with the wall the particle continues to have a negative velocity
(a successive collision will occur in the original model), the particle moves
beyond the wall. The modulus for the velocity injects the particle back and
fixes the problem.

\begin{figure}[htb]
\begin{center}
 \includegraphics*[width=6.0cm,angle=0]{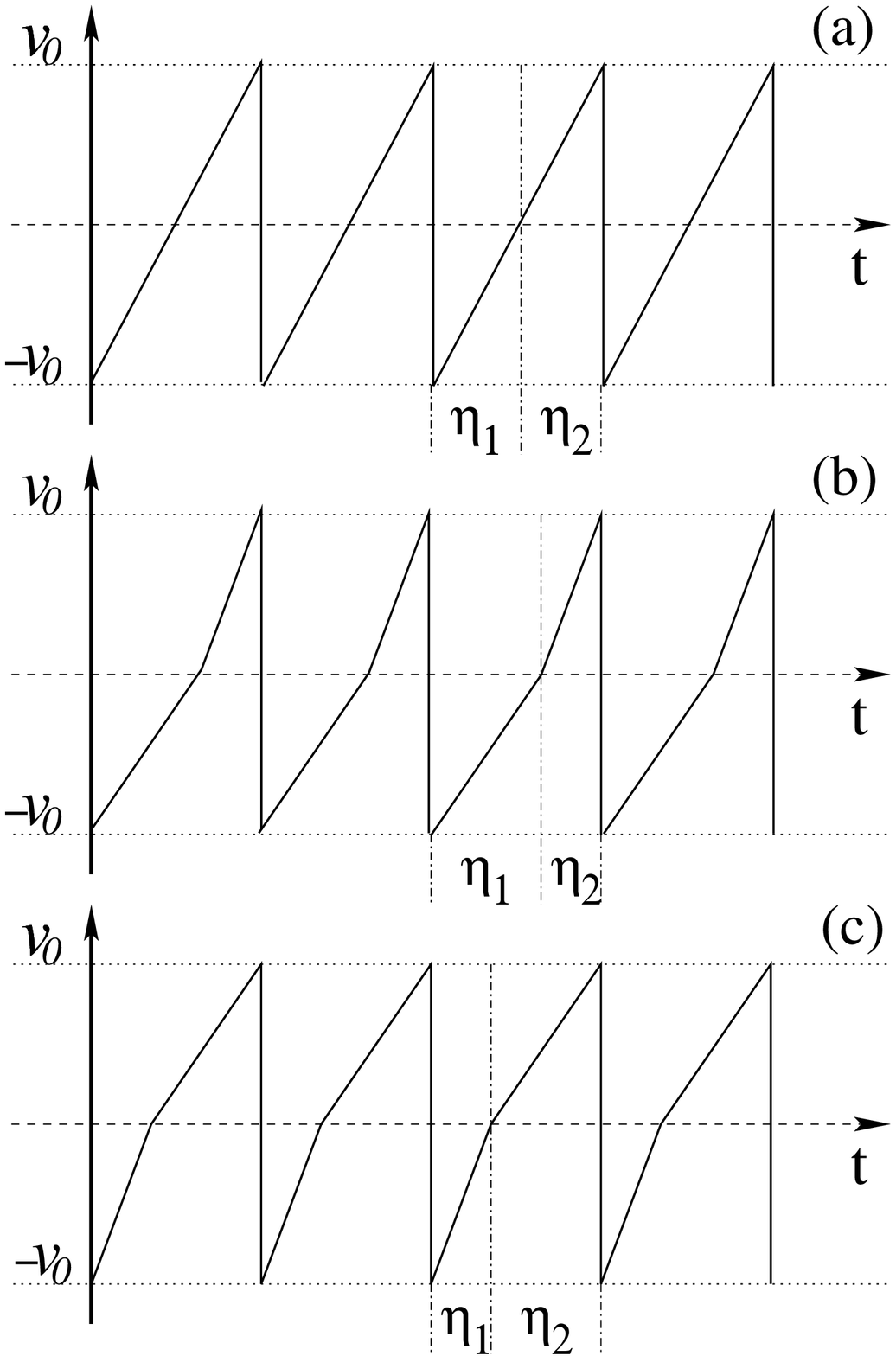}
\hspace*{2cm}
 \includegraphics*[width=5.0cm,angle=0]{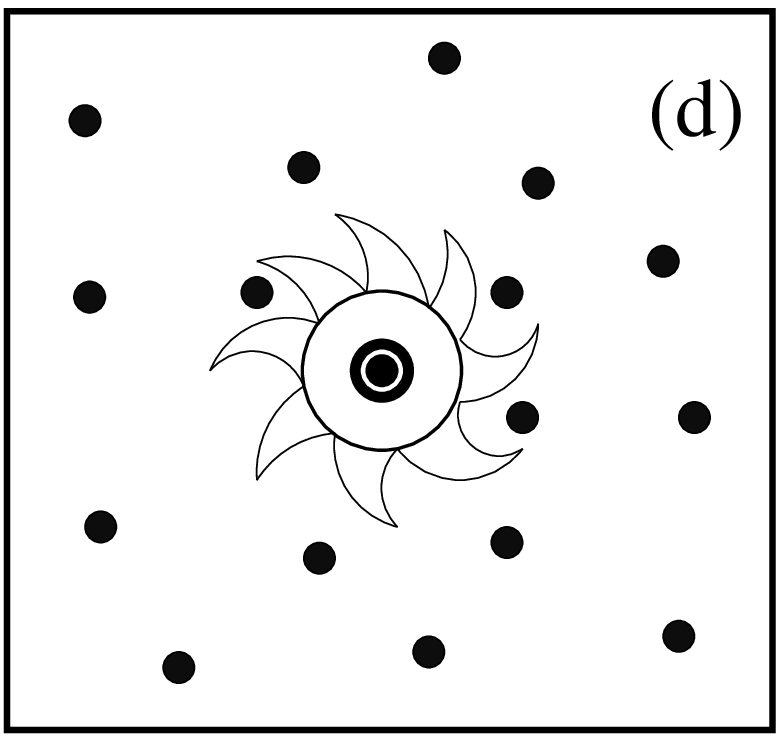}
\end{center}
\caption{\sl (a)-(c) The shape of the pulses used in the simulations. The 
deformed sawtooth effect is obtained when $\Delta\eta=\eta_2-\eta_1\ne0.0$. 
For (a) $\Delta\eta=0$ we have the symmetric {\it accelerated} case ${\bf A}$,
(b) $\Delta\eta<0$ we have the deformed {\it desaccelerated} case 
${\bf D}$ and (c) $\Delta\eta>0$ the deformed {\it hyperaccelerated} case 
${\bf H}$. (d) Illustration of a mechanical device which mimics the effect
of the soft version of the pulses from (b)-(c).  The ratchetlike cylinder 
may rotate clockwise and counter-clockwise.}
\label{bilhar}
\end{figure}

The time asymmetry of the oscillating wall in Eqs.~(\ref{map_1}) and
(\ref{map_2})  is controlled by varying the parameters ($\eta_1,\eta_2$).
The  {\it deformed} sawtooth (Ratchetlike) is obtained when
$\eta_1\ne\eta_2$. Figures \ref{bilhar}(a)-(c)
show the time behavior of the oscillating wall (the pulse) for different 
values of the asymmetry $\Delta\eta=\eta_2-\eta_1$. The deformed sawtooth
pulse is obtained when $\Delta\eta\ne0.0$. Such pulses can be easily obtained 
from pulse generators.  As we will see we obtain three kind of accelerations:
For $\Delta\eta=0$ we have the symmetric {\it accelerated} case ${\bf A}$
[Fig.~\ref{bilhar}(a)];  For $\Delta\eta<0$ we have the deformed {\it decelerated} 
case ${\bf D}$ [Fig.~\ref{bilhar}(b)]; and for $\Delta\eta>0$ the deformed 
{\it hyperaccelerated} case ${\bf H}$ [Fig.~\ref{bilhar}(c)]. Figure 
\ref{bilhar}(d) shows a very simple example of a {\it mechanical}
device which generates a similar effect as the deformed sawtooth. Imagine 
many particles colliding elastically with the walls inside the square box, and 
a ratchet cylinder rotating at the center. The rotating ratchet cylinder 
mimics the wall oscillation of the deformed sawtooth from
Fig.~\ref{bilhar}(b)-(c).  When an external motor starts to rotate the ratchet 
cylinder counter-clockwise, particles are hyperaccelerated. After that the
motor starts to rotate clockwise and the hyperaccelerated particles are decelerated. 
A similar mechanical device was used for micromotors in a bacterial
bath \cite{angelani09}.

\begin{figure}[htb]
\begin{center}
 \includegraphics*[width=6.0cm,angle=0]{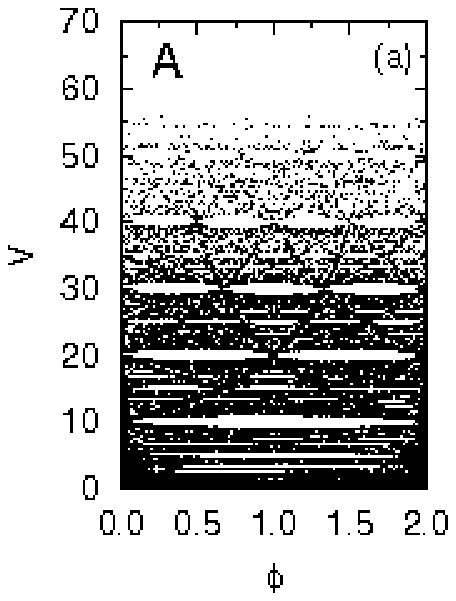}
 \includegraphics*[width=6.0cm,angle=0]{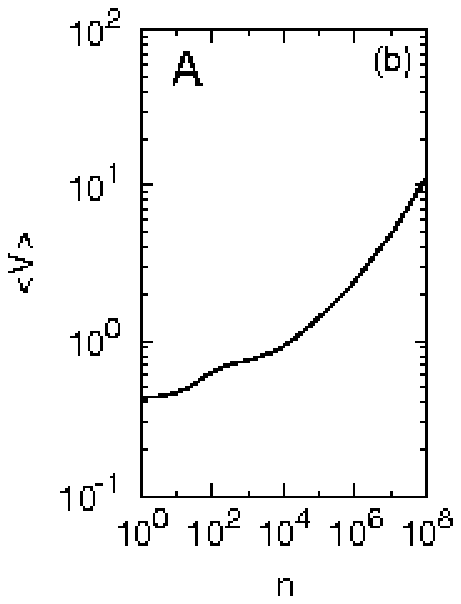}
 \includegraphics*[width=6.0cm,angle=0]{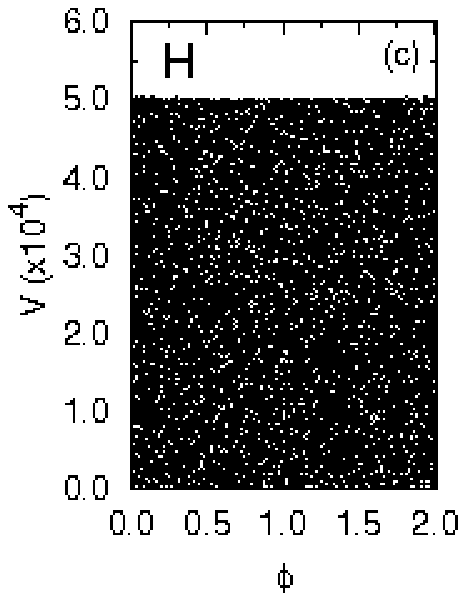}
 \includegraphics*[width=6.0cm,angle=0]{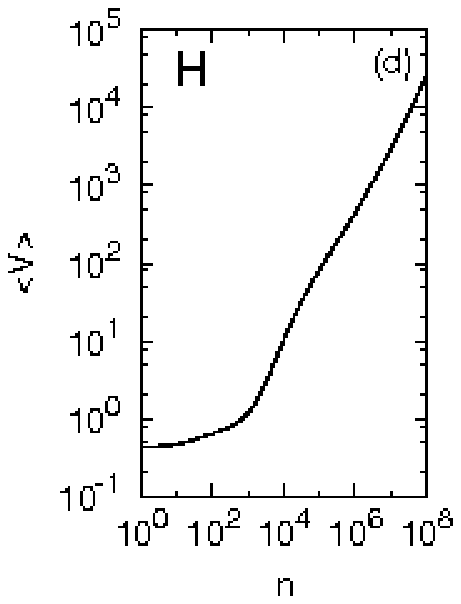}
 \includegraphics*[width=6.0cm,angle=0]{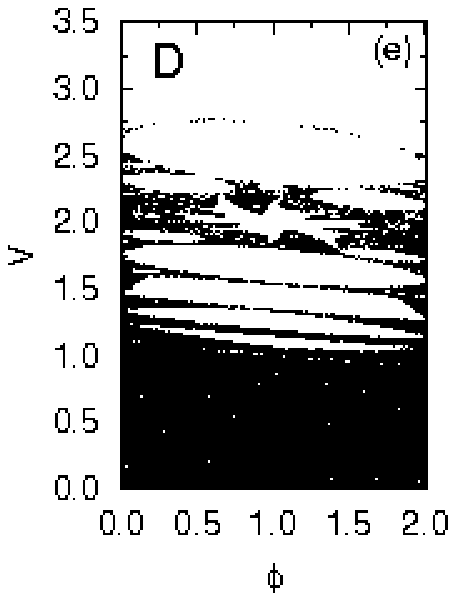}
 \includegraphics*[width=6.0cm,angle=0]{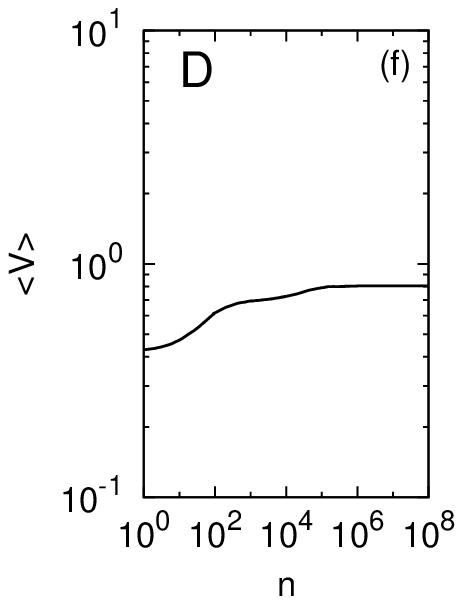}
\end{center}
\caption{\sl Evolution of $50$ chaotic orbits on the phase space
 $V\times\phi$ for the parameters $\mu=10$, $v_0=0.2$ and:
   (a) case ${\bf A:} \Delta\eta=0.0$, (c)~${\bf H:} \Delta\eta=0.01$ and
   (e)~${\bf D:} \Delta\eta=-0.01$. Mean values of the velocity (b), (d) and  
   (f) calculated over $3000$ trajectories related with the three case (a), (c)
   and (e), respectively. We call to attention that the order of magnitude from
 $\left<V\right>$ changes drastically between cases  ${\bf A,H,D}$.}
\label{velocity}
\end{figure}

Numerically we reckon the average particles velocity at a given time
$n$ from 

\begin{equation}
 \langle V \rangle(n) = 
       \frac{1}{n+1}\sum_{i=0}^{n}\frac{1}{\xi}\sum_{j=1}^{\xi}V_{n,j},
\end{equation}
where the index $i$ refers to the $i$th iteration of the sample $j$, and
$\xi$ is the number of initial conditions.
The average velocity $\langle V \rangle$ for the three cases ${\bf A,H,D}$ 
mentioned above are shown in Fig.~\ref{velocity}. We iterate the 
map~(\ref{map_1}) or (\ref{map_2}) for times $n=1\times 10^8$, and $3000$ 
initial conditions in the interval $0<V\le10^{-3}$ and 
$0\le\phi\le(\eta_1+\eta_2)$. Figure \ref{velocity} shows the phase space (on the 
left) and mean velocity (on the right) for the cases ${\bf A:} \Delta\eta=0.0$ 
in Figs.~\ref{velocity}(a)-(b); ${\bf H:} \Delta\eta=0.01$ in 
Figs.~\ref{velocity}(c)-(d); and
${\bf D:} \Delta\eta=-0.01$ in Figs.~\ref{velocity}(e)-(f). 
First observation is that the accelerated mode can be observed in the case
${\bf A}$. As the particle velocity increases, regular island are observed 
and $\langle V \rangle$ increases slowly until $\sim 10$. These regular islands 
prevent the particle velocity to increase very fast. It is worth to mention
that all initial conditions start inside the chaotic region at low velocities. 
The growth rate of $\langle V \rangle$ depends on the number of regular 
islands inside the phase space. This is an expected results since this case is 
similar to the simplified model studied by Lichtenberg and 
Lieberman~\cite{lichtenberg72,lichtenberg92}, where the harmonic force was 
considered. An interesting behavior appears when the deformed sawtooth
is introduced. For a very small asymmetry $\Delta\eta=0.01$ (case ${\bf H}$), 
the phase space is totally filled and no regular islands are observed 
[Fig.~\ref{velocity}(c)]. The corresponding $\langle V \rangle$ increases 
very fast until $\langle V \rangle\sim 2\times 10^{4}$ [see 
Fig.~\ref{velocity}(d)], showing that the accelerated mode 
is enhanced (hyperaccelerated) when compared to the case ${\bf A}$.
Another kind of motion occurs when $\Delta\eta=-0.01$  (case ${\bf D}$) 
where many regular islands appear in phase space [see 
Fig.~\ref{velocity}(e)] which again prevent the acceleration to increase without
bounds, as in case ${\bf A}$. The corresponding $\langle V \rangle$ 
remains almost constant for this example [Fig.~\ref{velocity}(f)]. A relevant
property of case ${\bf D}$ is that: if the initial condition for the velocity
is very high, the particle is decelerated. In other words, the hyperaccelerated 
particles from case ${\bf D}$ start to decelerate up to the time they reach the 
regular islands from Fig.~\ref{velocity}(e), but now from 
above. Therefore, there is a regular island barrier which prevents particles 
to decelerated until zero for short times. For very long times, however,
particles may pass between the regular islands and decelerate to zero
velocities. Such motion close to regular islands may lead to ``sticky'' or
trapped~\cite{zas02} trajectories which affect the velocity and the convergency 
of the finite time Lyapunov exponents~\cite{cesar1,cesar2,lopes05,cesarPRE09}.

In order to explain the effect of the deformed sawtooth pulse, we rewrite
maps (\ref{map_1}) and (\ref{map_2}) in the very high velocity regime
($V_0\gg1$). In this case the time evolution for the phase can be neglected 
and we have

\begin{equation}
 V_{n+1} \approx  nV_{0}\left(1+\frac{\Delta\eta\vert 
 \phi_0-\eta_1\vert}{\eta_1\eta_2} \right).
\label{map_r}
\end{equation}
From this expression for the velocity, we clearly see that when 
$\Delta\eta<0$ the velocity is reduced by each iteration and the decelerated 
mode is obtained. When $\Delta\eta=0$ the velocity is constant and when 
$\Delta\eta>0$ the velocity increases without bounds and we get the 
hyperaccelerated case.

\begin{figure}[htb]
\begin{center}
 \includegraphics*[width=8cm,angle=0]{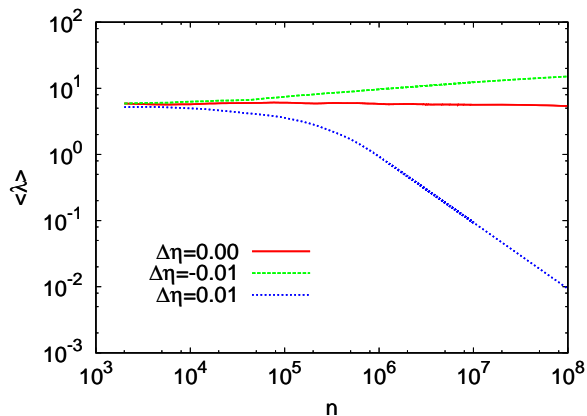}
\end{center}
    \caption{Mean value of the Lyapunov exponent calculated over 1000
      trajectories for the three situations: ${\bf A:}$ $\Delta\eta=0.0$ (full 
      line), ${\bf D:}$ $\Delta\eta<0.0$ (dashed line) and 
      ${\bf H:}$  $\Delta\eta>0.0$ (long-dashed line).}  
    \label{lyapunov}
\end{figure}

The above illustration shows the importance of a small deformation in
the sawtooth wall motion on the dynamics in phase space and on the corresponding 
acceleration. Now we look at the stability of the trajectories for each 
situation ${\bf A,H,D}$.
Figure \ref{lyapunov} shows the time evolution of the mean finite 
time Lyapunov exponent (LE), $\langle\lambda\rangle$, calculated over $1000$ 
trajectories for the three distinct cases ${\bf A,H,D}$. The LEs are determined
using the Benettin's method \cite{benettin80}. The full line shows  $\langle\lambda\rangle$
for the case ${\bf A}$, the long-dashed line for the case ${\bf H}$ and the 
dashed line for ${\bf D}$. In the accelerate case ${\bf A}$, 
$\langle \lambda \rangle$ is constant since every initial condition
(with low velocity) is trapped inside the chaotic region of phase
space. For the case ${\bf D}$, the $\langle \lambda \rangle$ is almost 
constant, increasing very slowly. The time behavior of the mean LE in case 
${\bf D}$ is similar to the mean LE from the ${\bf A}$ case in the following
sense: the trajectory  is trapped in the specific chaotic region in the phase
space, below the islands of regularity [please compare Figs.~\ref{velocity}(a) 
and (e)]. For the hyperaccelerated case ${\bf H}$ (long-dashed line) the mean LE 
decreases almost to
zero, showing that the filled space phase observed in Fig.~\ref{velocity}(c) is 
not chaotic. Albeit being not chaotic, no regular islands are observed, as in 
the other cases. This is apparently a typical behavior of linear unstable but 
ergodic systems \cite{casati76}, where besides having zero LEs the velocity
space is totally filled.

\begin{figure}[htb]
\begin{center}
 \includegraphics*[width=8cm,angle=0]{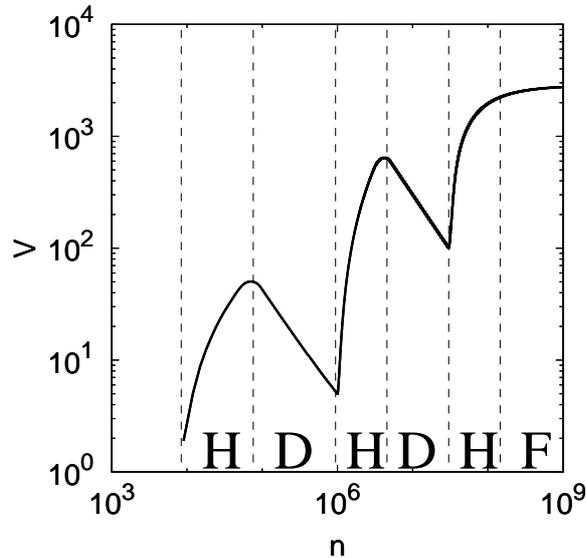}
\end{center}
    \caption{Velocity as a function of the iteration $n$. It shows how the
     particles velocity can be controlled by switching the sawtooth pulse 
      appropriately between cases ${\bf H}$ (hyperaccelerated) and ${\bf D}$
      (decelerated), and finally ${\bf F}$ (fixed wall).}   
    \label{control}
\end{figure}
At next we show how the Ratchetlike time asymmetry can be used to control the
Fermi acceleration in order to get a desirable velocity. Figure \ref{control} 
shows the time evolution of the particle velocity using a combination of
cases ${\bf H,D}$ discussed above. While for the case ${\bf H}$ the 
hyperaccelerated mode is obtained, the case ${\bf D}$ is able to 
decelerated the particle velocity. When the desirable velocity is reached, 
the wall oscillation can be turned off (case ${\bf F}$). By appropriately 
switching the Ratchetlike pulse between ${\bf H,D}$ modes, the velocity can
be controlled. It is important to mention that the control is only possible
due to the following particular property of the decelerated mode: the
deceleration only occurs after the particle is hyperaccelerated, it does not 
occur for small velocities [see Fig.~\ref{velocity}(e)]).

To conclude, for a long time the Fermi acceleration has been studied in 
different models and applications
\cite{fermi49,ulam61,lichtenberg72,lichtenberg92,leonel04,leonel04-prl,
leonel05,egydio06,karlis06,karlis07,pustylnikov77,steane95,schleich98,
saif05,milovanov01,michalek99,veltri04,kobayakawa02,malkov98,cole06}.
However, none of these works (but one \cite{karlis07}) considered the
possibility to control the particles in order to achieve a desirable
velocity. Here we show the effect of a {\it deformed} sawtooth (Ratchetlike) 
pulse to control the particle velocity in the Fermi-Ulam model. 
Changing the asymmetry parameter from the Ratchetlike pulse we are able to get
Fermi hyperacceleration and deceleration.  In fact, the time integral of the 
pulse (the wall position) over one period of oscillation must be different 
from zero in order to get the control. When the mentioned integral is positive 
(negative), the Fermi hyperacceleration (deceleration) occurs. This allows us 
to, by switching the pulse between hyperacceleration and deceleration modes, 
to control the particles velocity. We also show that the decelerated and normal
accelerated cases are chaotic, while for the hyperaccelerated case the LEs 
converge to zero. Additionally we give numerical evidences which indicate that 
the hyperaccelerated case is ergodic in velocity space. It would be interesting
to observe the deformed sawtooth pulse to control the particle velocity in the 
presence of dissipation \cite{cesarENG09}.

\section{Acknowledgments}
The authors thank CNPq and FINEP (under project CTINFRA-1) for financial
support.

%
\end{document}